\documentclass[runningheads]{llncs}
\usepackage{graphicx}
\usepackage{xspace}
\usepackage{color}
\usepackage{algorithm}

\usepackage{algpseudocode}
\usepackage{cite} 
\usepackage{booktabs}
\usepackage{appendix}
\usepackage[misc,geometry]{ifsym}
\newcommand{\system}{{JWBinder}\xspace}

\author{Yifan Xia\inst{1,2}, Ping He\inst{1}, Xuhong Zhang\inst{1}, Peiyu Liu\inst{1,2}, \linebreak  Shouling Ji\inst{1} , Wenhai Wang\inst{1,2}}
\begin{document}
\title{Static Semantics Reconstruction for Enhancing JavaScript-WebAssembly \linebreak Multilingual Malware Detection}
\vspace{-2cm}
\authorrunning{Y. Xia et al.}
\vspace{-0.2cm}
\institute{Zhejiang University, Zhejiang University NGICS Platform \linebreak\email{
        \{yfxia, gnip, zhangxuhong, liupeiyu, sji, zdzzlab\}@zju.edu.cn}
    }
\titlerunning{Static Semantics Reconstruction for Enhancing JWMM Detection}
\maketitle              % typeset the header of the contribution
\vspace{-0.6cm}
\begin{abstract}
The emergence of WebAssembly allows attackers to hide the malicious functionalities of JavaScript malware in cross-language interoperations, termed JavaScript-WebAssembly multilingual malware (JWMM).
However, existing anti-virus solutions based on static program analysis are still limited to monolingual code. 
As a result, their detection effectiveness decreases significantly against JWMM. 
The detection of JWMM is challenging due to the complex interoperations and semantic diversity between JavaScript and WebAssembly.
To bridge this gap, we present \system, the first technique aimed at enhancing the static detection of JWMM.
\system performs a language-specific data-flow analysis to capture the cross-language interoperations and then characterizes the functionalities of JWMM through a unified high-level structure called Inter-language Program Dependency Graph.
The extensive evaluation on one of the most representative real-world anti-virus platforms, VirusTotal, shows that \system effectively enhances anti-virus systems from various vendors and increases the overall successful detection rate against JWMM from 49.1\% to 86.2\%. 
Additionally, we assess the side effects and runtime overhead of \system, corroborating its practical viability in real-world applications.
\vspace{-0.2cm}
\keywords{Malware and Unwanted Software \and Software Security \and Web Security.}
\end{abstract}
\vspace{-1cm}
\section{Introduction}
\label{sec:intro}
%1. js 语言的重要性
\vspace{-0.2cm}
JavaScript is a highly prevalent scripting language known for its significant role in web application development \cite{languageranks}. In recent years, it has also extended its influence beyond the browser with the support of NodeJS\cite{nodejs}. 
The ubiquity of JavaScript naturally makes it a target for attackers, giving rise to a variety of attack vectors, such as CryptoJacking \cite{cryptojacking}, Drive-by-download attacks\cite{drivebydownload} and JavaScript Skimmers\cite{skimmer}. 
Additionally, attackers have now started exploiting Open Source Software (OSS) by injecting malicious JavaScript third-party packages into public registries like NPM\cite{npm}.

%2. 现有的检测方法
To counter these threats, current anti-virus solutions employ 
sophisticated program analysis techniques for malicious JavaScript detection \cite{cujo, zozzle, rozzle, JSDC, jstap, doublex, yara, Hiddenode, js*, boxjs, de4js, maljail}. Such approaches can be partitioned into two categories: static and dynamic approaches. 
Static approaches extract code features of varying granularities (e.g., Abstraction Syntax Tree (AST)\cite{zozzle} and Program Dependence Graph (PDG)\cite{jstap}) from JavaScript without executing it. 
These features are then used for machine learning techniques \cite{cujo, zozzle, jstap} or program similarity analysis \cite{JSDC, js*} to differentiate benign and malicious code. 
On the other hand, dynamic approaches detect abnormal JavaScript behavior (e.g., sensitive API calls) by running it in a honey client or sandbox \cite{boxjs,maljail}. 
Each approach has its own strengths and weaknesses. 
However, dynamic approaches are often burdened with considerable runtime overhead and struggle to detect malicious behaviors only manifesting under specific configurations.
Thus, static approaches often form the preferred choice for anti-virus solutions due to their scalability and efficiency, consequently making them a prime target for attackers \cite{hidenoseek, wobfuscator, evadestatic}.

% 3. 已有的检测方法面对跨语言malware的weakness
Despite the considerable ability of static approaches to detect malware, existing defense methods tend to assume that programs in the JavaScript ecosystem (e.g., Web and NodeJS) are composed purely of JavaScript. 
However, this assumption may no longer be held with the introduction of WebAssembly \cite{wasm} in 2015. 
WebAssembly is an emerging binary code language that complements JavaScript. 
Initially designed for computation-intensive tasks, WebAssembly can be called upon by JavaScript programs through foreign language interfaces, which also provides new opportunities for attackers to create JavaScript-\linebreak WebAssembly Multilingual Malware (JWMM) \cite{wobfuscator, mwasm}. 
Specifically, attackers can conceal malicious behaviors within the interoperations between JavaScript and WebAssembly.
Consequently, prior works that statically extract program features solely from JavaScript \cite{cujo, zozzle, rozzle, JSDC, jstap, doublex, yara, Hiddenode, js*} could hardly identify these concealed malicious behaviors.
Even the existing works \cite{romano2020minerray} that consider malicious WebAssembly struggle to mitigate this threat because the malicious behaviors of JWMM are concealed behind the cross-language interoperations, which are unlikely to be identified with detectors focusing on a single language.
% 4. 讲述难点和解决方案
% 难点一：捕捉跨语言交互
% 难点二：将针对js malware设计的检测方法拓展到跨语言malware

The effectiveness of JWMM against static approaches raises legitimate concerns. 
However, the detection of JWMM presents two major challenges:

% \vspace{-0.6cm}
C1: \textit{Interoperation Complexity.} 
A fundamental step in characterizing JWMM involves understanding how JavaScript interacts with WebAssembly units in JWMM. 
This is far from a trivial task due to the complexity stemming from the intricate mechanisms upon which the interoperations of JavaScript and WebAssembly depend. 
For example, there are various interfaces to initialize a WebAssembly instance in JavaScript. 
Additionally, both languages can import and invoke functions from each other in adherence to the language standard \cite{mdn}.
Without recognizing these interoperations, it is difficult to unify the semantics of different language units in JWMM.
% \vspace{-0.2cm}

C2: \textit{Semantics Diversity.} 
Existing works capture various patterns and features in sole language for distinguishing monolingual malware \cite{doublex, yara,js*, romano2020minerray}.
However, JavaScript and WebAssembly have disparate language semantics.
Therefore, the characterization of the JWMM's functionalities necessitates the consideration of both JavaScript and WebAssembly semantics.
Furthermore, even if we consider the semantics of both JavaScript and WebAssembly, the definition of malicious patterns/features in multilingual programs remains an open issue.

In this paper, we present \system, the first technique that characterizes the functionalities of JavaScript-WebAssembly multilingual programs to enhance the static detection of JWMM.
To tackle C1, %the interoperation complexity challenge, 
\system presents a language-specific data-flow analysis to capture the interoperations between JavaScript and WebAssembly.
Specifically, for the target JWMM, \system constructs and traverses the PDG of its JavaScript unit. 
Then, by tracing the data dependencies flowing into and out of the foreign language interfaces, \system can identify the concealed WebAssembly units in JWMM and detect how JavaScript and WebAssembly interact with each other. 
For example, the instantiation of WebAssembly which passes JavaScript external functions to WebAssembly instances, or the invocation point of a WebAssembly internal function in the JavaScript unit.

The solution of C2 %the semantics diversity issue 
relies on the crucial observation that, 
while WebAssembly's semantics differs significantly from JavaScript, it shares some unified features derived from JavaScript, such as similar control-flow instructions and basic data instructions \cite{wasmins}. 
Furthermore, WebAssembly can only \textcolor{black}{invoke privileged system functions} by importing them from JavaScript rather than through customized implementation. 
These homogeneous features allow us to design a uniform abstract representation that characterizes the functionalities of JavaScript-WebAssembly multilingual programs at a high level.

Based on the above insight, we propose a novel technique called \textit{Static Semantics Reconstruction} (SSR). 
\textcolor{black}{Leveraging the homogeneous features between JavaScript and WebAssembly, we first design a set of abstraction rules which encapsulate the semantics of WebAssembly at a high level.}
Following our abstraction rules, SSR generates a JavaScript-like abstract representation of the JWMM's WebAssembly units. 
Then, SSR integrates the abstract representations of WebAssembly units into JavaScript PDG to construct a uniform structure termed Inter-language PDG (IPDG), which characterizes the semantics within and across the JavaScript and WebAssembly units.
Rather than designing ad-hoc heuristics to enumerate the malicious patterns of JWMM using the IPDG, 
SSR's final phase involves translating the IPDG and reconstructing a pure JavaScript program, which serves to elucidate the functionalities of the initial JWMM. 
The rationale is that since the IPDG is originally constructed following JavaScript abstract syntax, it can be naturally translated back to JavaScript by node traversing. Anti-virus solutions can further examine this reconstructed pure JavaScript program to determine the malignancy of the original JWMM.

The reconstruction of pure JavaScript programs offers two primary benefits. 
Firstly, it transforms the challenge of identifying multilingual malware into the more straightforward task of detecting monolingual malware. 
This allows the reuse of detection patterns/features designed for monolingual malware, thereby effectively addressing the semantic diversity problem. 
Secondly, with \system acting as a preliminary process, existing anti-virus solutions that concentrate solely on JavaScript can be employed directly to detect JWMM without any modifications, enhancing this approach's practicality for real-world applications.

% 6. 实验结果
To validate our design, we build a JWMM dataset based on 44,369 real-world JavaScript malware and evaluate \system using this dataset. 
We have the following results.   
First, the approach is effective in enhancing the detection capabilities of well-known commercial Anti-Virus Systems (AV-Systems):
With the application of \system, the overall successful detection rate of VirusTotal\cite{virustotal} increases from 49.1\% to 86.2\%. 
Meanwhile, the number of AV-Systems successfully detecting malicious samples has increased from 4.1 to 8.3 on average. 
Second, \system introduces nearly no side effects to benign programs:
Processing a benign JavaScript-WebAssembly multilingual programs dataset and uploading them to VirusTotal, the results show minimal differences (0.5\% false positive rate) compared to the original benign cases.
This demonstrates that \system does not induce suspicious behaviors which influence the detection of benign multilingual programs.
Third, we investigate the generalization ability of \system on AV-Systems from different vendors. The results show that more than 10 different AV-Systems significantly benefit from \system. In particular, AV-Systems from different vendors may favor particular variants of \system.
Lastly, we find that our tool only requires, on average, 25.6 seconds to process a single JWMM program (with an average size of 282 KB), which is acceptable in comparison to previous JavaScript program analysis works \cite{doublex, odgen}.
Collectively, these results indicate that \system is a practical tool for the enhancement of real-world anti-virus solutions.
% 7. contribution

In summary, this paper offers the following contributions:
\begin{itemize}
\item We propose the first program analysis technique designed to counter JWMM, which solutes the challenge of interoperation complexity and semantics diversity for analyzing JWMM.
\item We implement the prototype of \system, with a data-flow analysis framework for capturing cross-language interoperations in JWMM and a novel method termed static semantic reconstruction to characterize the unified functionalities of JWMM.
\item We conduct a comprehensive evaluation demonstrating that our approach effectively enhances state-of-the-art anti-virus solutions and provides an analysis of their internal mechanisms.
\end{itemize}

To foster further research, we will release our experiment data and implementation at \cite{jwbinder}. We believe that \system provides valuable insights to detect JavaScript-WebAssembly multilingual malware. 
\vspace{-0.3cm}
\section{Backgorund and Motivation}
\label{sec:background}
\vspace{-0.1cm}
% WebAssembly
\subsection{WebAssembly}
WebAssembly\cite{wasm}, a low-level binary instruction format, has become a fundamental web standard due to its secure and efficient characteristics. 
It facilitates web development by enhancing the performance of applications and enabling seamless integration with JavaScript.

The WebAssembly binary includes multiple sections. Most notably, its code section holds the functional components, while the memory and data section manages linear memory for runtime behavior. 
The instructions of WebAssembly work at a low level to comprise simple operations such as arithmetic, control flow, and memory access. 
Some of the instructions share similar semantics to high-level languages (e.g., loop, branch, variable declaration/usage instructions), while others have special functionality corresponding to specific features (e.g., memory instructions, bit-level numeric instructions) of WebAssembly. 

WebAssembly does not have a standard library, which means it can not directly access system APIs. 
To perform such functionalities, WebAssembly must import external functions from its host language (e.g., JavaScript). 
To achieve this, JavaScript uses foreign language interfaces (FFI) \cite{mdn} to modularize and instantiate (that is, fulfill the imports of) WebAssembly modules and call exported WebAssembly functions.
% JWMM
\vspace{-0.4cm}
\subsection{JavaScript-WebAssembly Multilingual Malware} 
We define JavaScript-WebAssembly Multilingual Malware (JWMM) as a family of malware which hides malicious behaviors across the interoperations between JavaScript and WebAssembly.  
To evade the detection of anti-virus solutions, known JWMM often abuses WebAssembly binary for hiding sensitive instructions and data, and changing the control flow of JavaScript\cite{mwasm, wobfuscator}. 

Alan et al.\cite{wobfuscator} developed Wobfuscator to generate JWMM automatically and demonstrated its efficacy against academic machine-learning-based classifiers. In this paper, we extend this investigation by testing JWMM's evasion abilities against a leading anti-virus platform, VirusTotal. We found that over half of the JWMM samples successfully evaded VirusTotal's detection, more details will be discussed in Section \ref{sec:evaluation}.

% Motivation Example
\vspace{-0.5cm}
\subsection{A Motivating Example}

Figure 1 depicts a motivating example to illustrate how JWMM evades the detection of existing anti-virus solutions for JavaScript malware. 

\begin{figure}[]
    \centering
    \includegraphics[width=12cm]{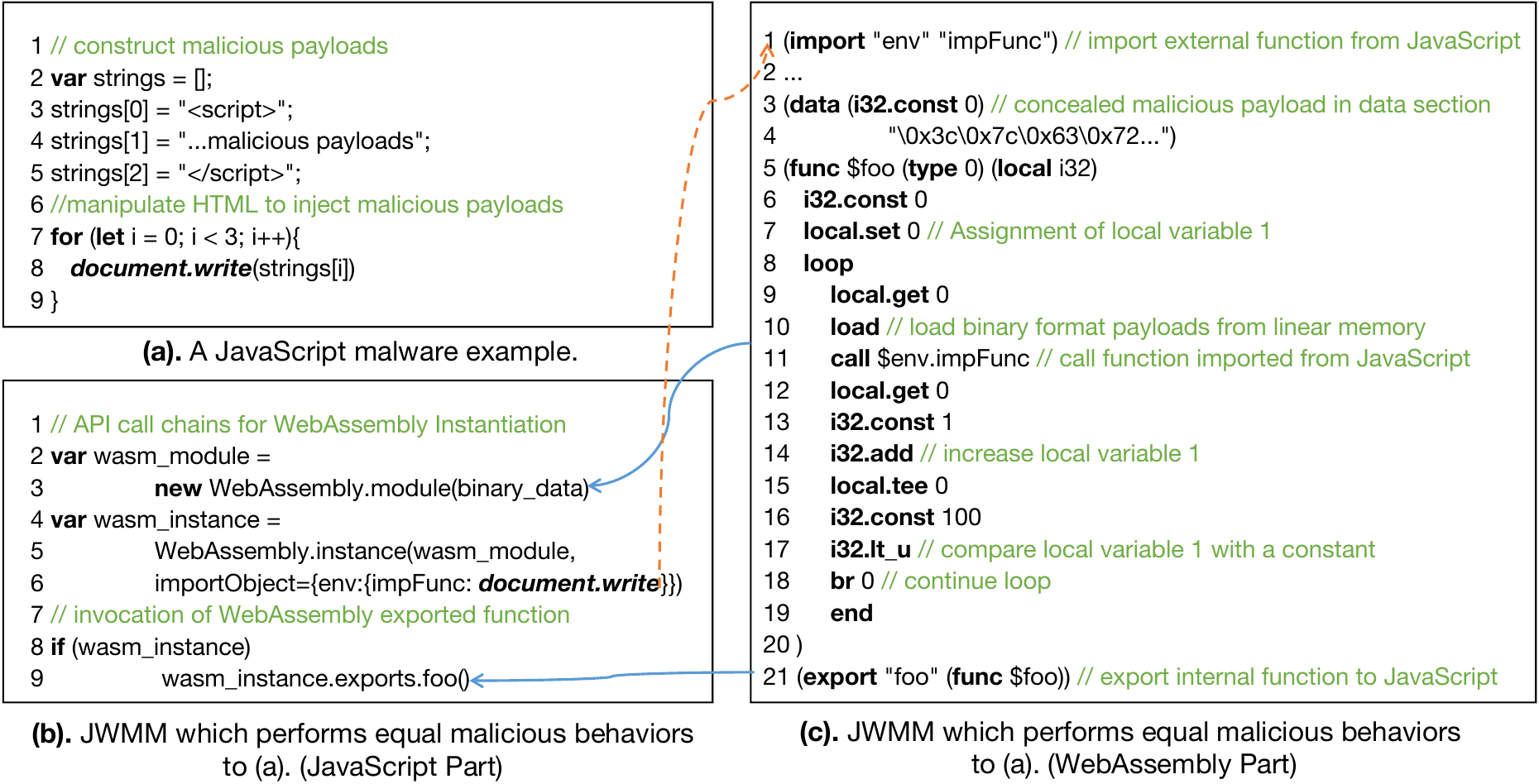}
    \caption{An example of JWMM (b and c) and its equivalent JavaScript malware (a)}
    \label{fig:motivation}
    \vspace{-0.5cm}
\end{figure}

Figure \ref{fig:motivation}.a shows a real-world pure JavaScript malware successfully detected by McAfee-GW-Edition\cite{mcafee}. 
To conduct the attack, it iteratively executes the \textbf{document.write} function (lines 7 - 9) to insert pre-defined malicious payloads (lines 2 - 5) into HTML. 
Figure \ref{fig:motivation}.b and Figure \ref{fig:motivation}.c present the JavaScript part and WebAssembly part (in human-readable format) of the JWMM which perform equivalent malicious functionalities to Figure \ref{fig:motivation}.a. 

Next, we explain how this JWMM evades detection by abusing the interoperations between JavaScript and WebAssembly.
In Figure \ref{fig:motivation}.b, the JavaScript part of JWMM only needs to instantiate a WebAssembly instance by calling the FFI sequence (lines 2 - 6), and then invokes the internal function \textbf{foo} exported from the WebAssembly part (the blue solid line). 
In other words, it conceals the malicious functionalities in the WebAssembly.
While in Figure \ref{fig:motivation}.c, the WebAssembly part imports the \textbf{document.write} from JavaScript (the red dashed line) and iteratively executes the function in a loop (lines 8 - 19) with transformed binary-format malicious payloads stored in its data section (lines 3 - 4). 

Existing anti-virus solutions designed for pure JavaScript are not effective in detecting such JWMM.
Also, merely analyzing the WebAssembly binary additionally could hardly reveal cross-language malicious functionalities if we are not aware of the exact imported function from JavaScript (i.e. \textbf{document.write}). 

This example not only illustrates how JWMM evades the detection of monolingual anti-virus solutions, thereby emphasizing the necessity for a cross-language, comprehensive analysis, 
but also motivates our insights to reconstruct JWMM to a monolingual format, as depicted in Figure \ref{fig:motivation}.a. 
By doing so, we can expose its malicious functionalities for monolingual anti-virus solutions to identify.
 % and motivation
\vspace{-0.4cm}
\section{JWBinder}
\label{sec:methodology}
\vspace{-0.2cm}
This section describes our technique approach. 
We start with an overview (\S \ref{subsec:overview}) of \system and then elaborate two critical phases: language-specific data-flow analysis (\S \ref{subsec:dfa}) and static semantic reconstruction (\S \ref{subsec:ssr}).
\vspace{-0.4cm}
\subsection{Approach Overview}
\label{subsec:overview}

\begin{figure}[]
    \centering
    \includegraphics[width=1 \linewidth]{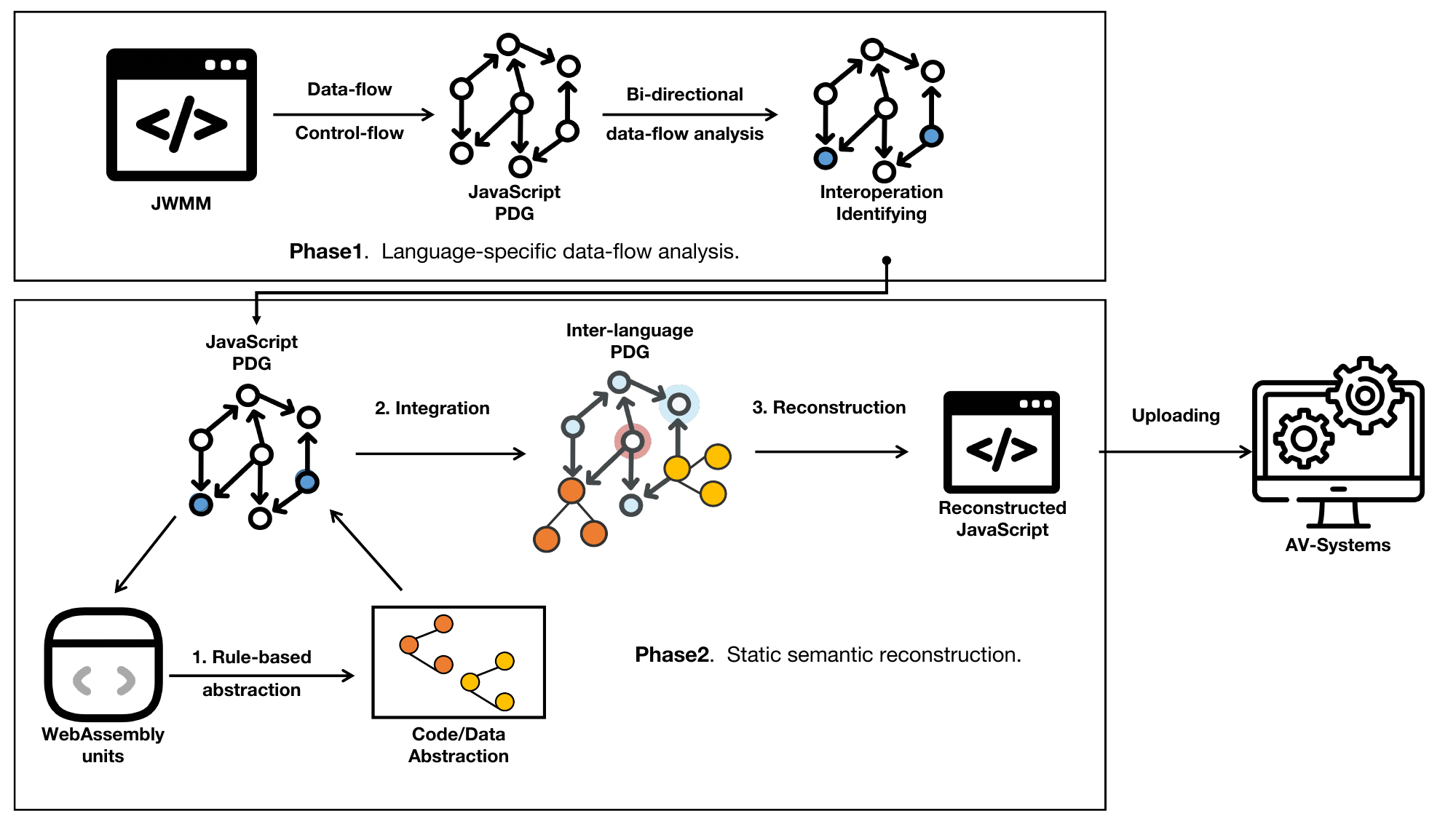}
    \caption{An overview of \system}
    \label{fig:overview}
    \vspace{-0.6cm}
\end{figure}

Figure \ref{fig:overview} gives the overview of \system. 
The input of \system is the \linebreak JavaScript-WebAssembly multilingual program under detection.
With the input, \system works in two phases.
In Phase 1, \system first abstracts the JavaScript unit of the multilingual program to construct the JavaScript PDG. 
We adopt the definition of PDG followed Fass et al. \cite{doublex}, which integrates the control-flow and data-flow dependencies into the Abstract Syntax Tree of the JavaScript unit.
Next, \system traverses the PDG to perform a bi-directional data-flow analysis which captures the interoperation between JavaScript and WebAssembly. 

After Phase 1, the functionalities of WebAssembly are still invisible on the JavaScript PDG. 
Thus, \system starts its Phase 2 to construct a language-agnostic structure for characterizing the functionalities of multilingual programs.
In Phase 2, \system first extracts the WebAssembly units in the multilingual program based on the identified interoperations in Phase 1.
For every individual WebAssembly unit, \system abstracts its code and data sections following a set of abstraction rules, which concentrate on the homogeneous semantics between JavaScript and WebAssembly.
The abstractions of WebAssembly units are then integrated into the JavaScript PDG to construct a uniform language-agnostic structure termed Inter-language Program Dependency Graph (IPDG). 
Finally, \system reconstructs a pure JavaScript program based on the IPDG, transforming the problem of detecting multilingual malware back into monolingual malware. 
As a result, JWBinder outputs pure JavaScript programs that statically characterize the original multilingual program's functionalities, making these outputs ready for detection by monolingual anti-virus solutions (e.g., AV-Systems).
\vspace{-0.2cm}
\subsection{Phase 1: Language-specific Data-flow Analysis}
\label{subsec:dfa}
At a high level, a JavaScript-WebAssembly multilingual program comprises a JavaScript program that interacts with WebAssembly through specific interfaces, with the WebAssembly units concealed in the JavaScript unit.
To characterize the functionalities of the multilingual program, \system should be able to recognize how JavaScript interacts with WebAssembly units. 
To this goal, the first phase works in two major steps as elaborated below.

\textbf{PDG Generation.}
JWBinder first abstracts the JavaScript unit of the multilingual program for an in-depth data-flow analysis. 
Given the JavaScript unit, \system generates its PDG following the definition of Fass et al. \cite{doublex}.
The PDG is built on the JavaScript AST, incorporating control-flow and data-flow dependencies during AST traversal.
In the PDG, The control-flow dependencies present a decision pathway, delineating whether a specific execution path would be pursued. 
Concurrently, the data-flow dependencies offer insights into the interrelationships among different variables within the program.

\begin{figure}
    \vspace{-0.5cm}
    \centering
    \includegraphics[width=\linewidth, height=4cm]{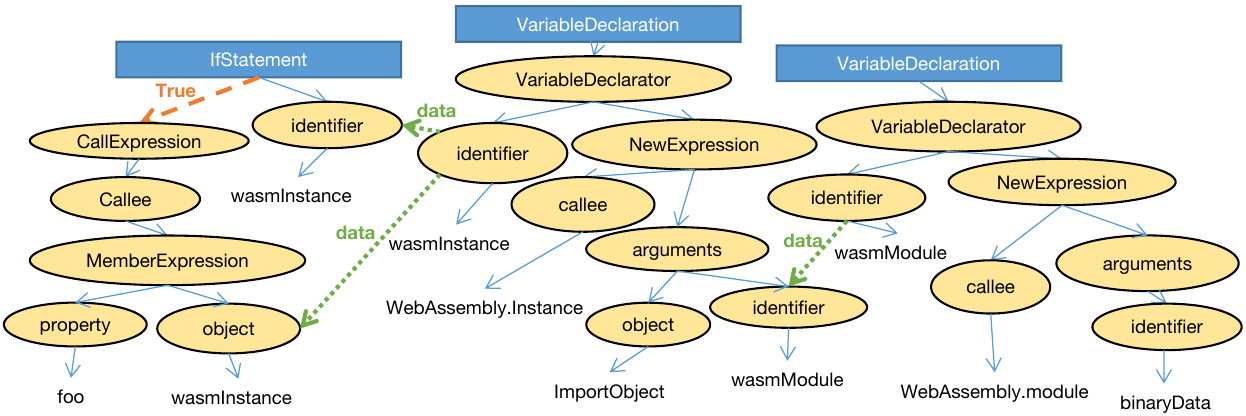}
    \caption{Example PDG of figure 1.b (some nodes are simplified for clarity)}
    \label{fig:pdg}
    \vspace{-0.4cm}
\end{figure}
For instance, Figure \ref{fig:pdg} is the PDG of the JavaScript program shown in Figure \ref{fig:motivation}.b. 
The control dependency labeled as ``True'' (red dashed line) suggests that the program path will be pursued only when the preceding condition holds true. 
Meanwhile, the data dependencies (green dot lines) present the definition and usage relationships of different variables (e.g., wasmInstance and wasmModule). 
Both these dependencies are valuable for identifying the interoperation between JavaScript and WebAssembly.

\textbf{Interoperation Identification.}
Once the JavaScipt PDG is constructed, 
\system conducts a bi-directional data-flow analysis to capture the interoperation between JavaScript and WebAssembly. 
The multilingual interoperation patterns normally contain the modularization/instantiation of WebAssembly and the invocation of WebAssembly properties. 
The former process enables JavaScript to pass external properties into WebAssembly, while the latter makes JavaScript invoke the internal properties of WebAssembly.
For example, in Figure 1.b, the JWMM compiles a WebAssembly module (\textit{wasmModule} in lines 2 - 3), employs the module to instantiate a WebAssembly instance (\textit{wasmInstance} in lines 4 - 6) with external function \textit{document.write()}, and finally calls an internal function from the WebAssembly instance (\textit{wasmInstance.foo()} in line 9). 

To capture the above interoperation patterns, \system traverses the PDG to identify the key APIs and properties (listed in Figure \ref{fig:keyfunc} within the Appendix) for WebAssmebly modularization or instantiation. 
Upon encountering one of these key APIs, JWBinder marks the relevant PDG nodes as interoperation positions and discerns the external properties passed into WebAssembly through the APIs. 
Then \system undertakes a forward data-flow analysis to trace the locations where JavaScript invokes internal properties of the WebAssembly instances.
Meanwhile, a backward data-flow analysis is also conducted to trace the concealed binary format WebAssembly for further abstractions in Phase 2. We detail the algorithm in 
Algorithm \ref{alg} within the Appendix.
Ultimately, \system generates a JavaScript PDG where the cross-language interoperations are distinctly marked, providing a clear depiction of the intricate connections between JavaScript and WebAssembly.
\vspace{-0.4cm}
\subsection{Phase 2: Static Semantic Reconstruction}
\label{subsec:ssr}
\vspace{-0.2cm}
\system performs the static semantic reconstruction (SSR) to characterize the functionalities of JWMM after the interoperations between JavaScript and WebAssembly have been identified. 
The key insight behind SSR is the integration of JavaScript and WebAssembly semantics into a uniform representation based on their homogeneous features. 
In particular, Phase 2 works in three steps as elaborated below.

\textbf{Rule-based Abstraction.} 
A joint analysis for multilingual programs requires a uniform representation that merges both JavaScript and WebAssembly semantics.
Since the main module of JWMM is the JavaScript unit, \system integrates the semantics of WebAssembly units into the JavaScript PDG generated in Phase 1 for multilingual analysis. 
However, this is a non-trivial process due to the disparate semantics of JavaScript and WebAssembly. Most importantly, JavaScript is a dynamically typed language, while WebAssembly is a statically typed language, which has diverse data types. 
The disparity makes it difficult to unify the cross-language semantics within JWMM.

\system addresses this challenge by introducing a set of abstraction rules, which extracts high-level semantics from WebAssembly to unify different language units.
The critical insight of our abstraction rules is that they focus on homogeneous features between JavaScript and WebAssembly, such as similar data-related operations (e.g., variable declaration and usage) and logic-related instructions (e.g., loop structures and condition statements). 
Based on this, \system is able to translate the semantics of WebAssembly to JavaScript-like AST nodes and bridge the semantic gap between JavaScript and WebAssembly.

In general, the abstraction rules fall into two categories: code abstraction and data abstraction, targeting code and data sections mentioned in \S \ref{sec:background}. 
We now introduce different categories of abstract rules.

The \textbf{code} abstraction rules aim to characterize the functionalities of internal functions defined in WebAssembly's code section. 
For example, in Figure \ref{fig:motivation}.b, the definition of the function \textit{foo} (line 9) lies in WebAssembly's code section. 
SSR reveals its functionality by abstracting the corresponding WebAssembly instructions (lines 5-20 in Fig \ref{fig:motivation}.c).
WebAssembly code instructions execute on a stack machine, in that instructions manipulate values on an implicit operand stack, consuming (popping) argument values and producing or returning (pushing) result values\cite{wasmins}. 
To abstract these instructions, we should first capture the data relationships between stack values.
Therefore, we model WebAssembly instructions based on their effects on the stack, following the simulation of \cite{romano2020minerray}. 
According to the stack state, we design specific rules for different code instructions to extract their high-level semantics and abstract them to JavaScript ES6 \cite{es6} syntax units. 
We choose ES6 syntax abstractions because it helps to characterize WebAssembly instructions within JavaScript syntax, which strengthens the homogeneity between these two languages.
\begin{table}[]
\vspace{-0.5cm}
    \centering
    \caption{Abstraction rules}
    \includegraphics[width=\linewidth]{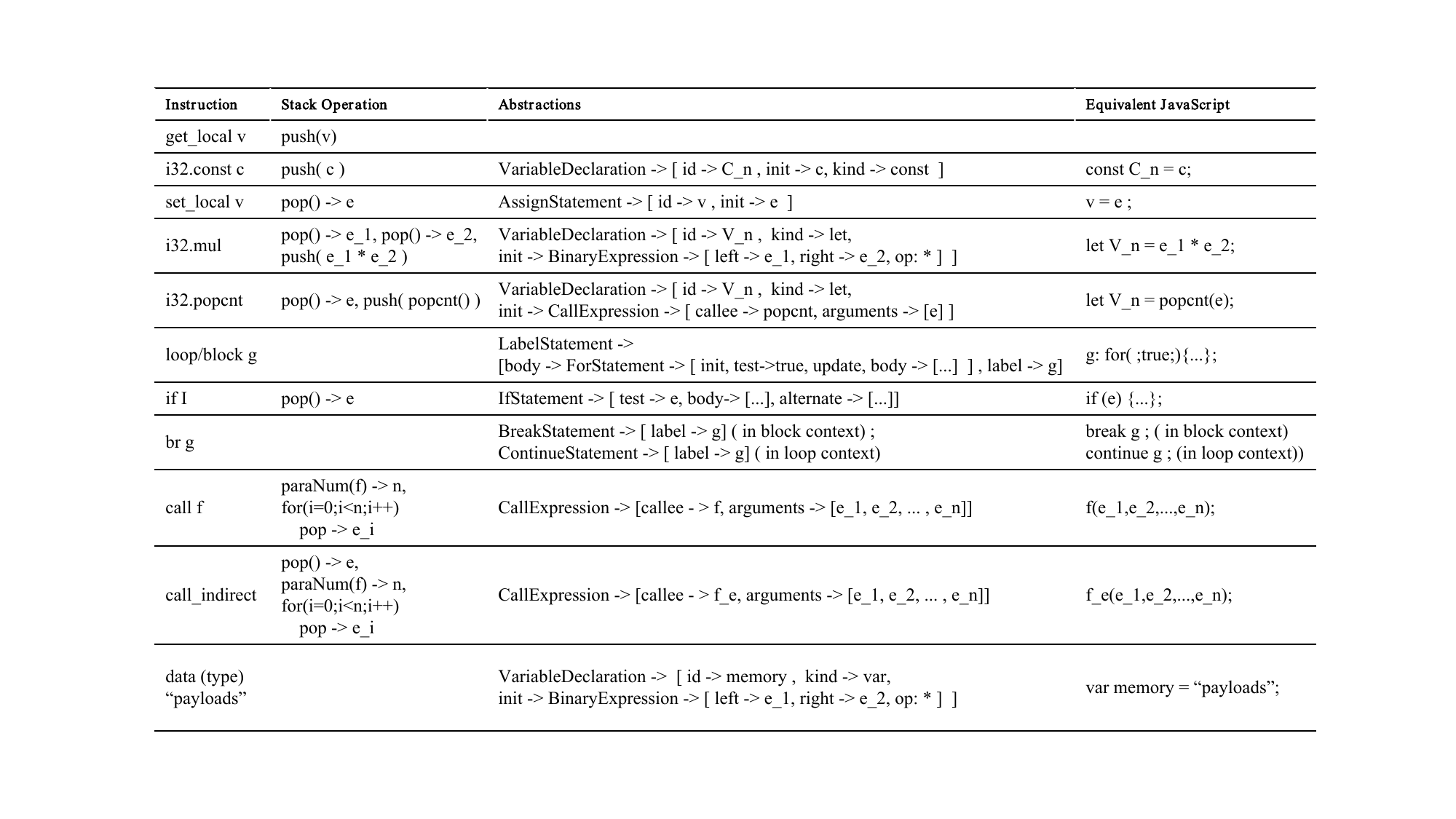}
    \label{fig:abstraction}
    \vspace{-1cm}
\end{table}

Table \ref{fig:abstraction} shows a subset of the abstraction rules. 
Due to the space limit, we present the full version on our website\cite{jwbinder} and only list abstraction rules for the key data-flow and control-flow instructions. 
For instructions that do not assign variables or manipulate control flows, such as ``local\_get'' which only fetches an existing local variable and pushes it onto the stack, we models their stack operation without generating abstractions. 

\textit{Data-flow instruction}. WebAssembly functions operate their variables following the ``def-use'' chain like high-level programming languages. We abstract such data-flow instructions as the foundation to characterize the semantics of WebAssembly's code section.
Specifically, we abstract instructions that manipulate local/global variables to ``AssignStatement'' syntax units. 
As shown in row 4, Table \ref{fig:abstraction}, \textit{``set\_local v''} consumes a stack value \textit{e} and assigns \textit{e} to local variable \textit{v}. 
The abstraction of \textit{``set\_local v''} indicates this instruction has equivalent functionality compared to JavaScript code \textit{``v = e''}.
Instructions that push new values to the stack are abstracted to ``VariableDeclaration'' syntax units: 
\textit{``i32.const c''} declares a new constant and pushes it onto the stack. 
Its abstraction is equivalent to \textit{``const C\_n = c;''}, where \textit{"C\_n"} is a randomly generated name for evading name conflicts. 
Similarly, the instruction \textit{``i32.mul''} and \textit{``i32.popcnt''} are abstracted to new variable declarations. In particular, operators that are unsupported by JavaScript are simulated to user-defined functions (e.g., \textit{popcnt} in row 5).
The rest of Table \ref{fig:abstraction} shows code abstraction rules for WebAssembly control flow instructions.

\textit{Control-flow instruction}. WebAssembly supports control flow instructions, which are similar to high-level languages such as JavaScript. 
These instructions are significant for constructing a uniform representation. 
As shown in Table \ref{fig:abstraction}, the instructions \textit{``loop''} and \textit{``block''} are both abstracted to labeled ``ForStatement'' syntax units in Table \ref{fig:abstraction}. 
However, they construct different contexts where the \textit{``br''} instruction is abstracted to different syntax units (i.e., Break and Continue), resulting in diverse control flow structures. 
\textit{``if I''} is abstracted to ``IfStatement'', which is a homogeneous structure in JavaScript. 
In particular, the \textit{``call''} and \textit{``call\_indirect''} instructions are non-deterministic because they can either call WebAssembly internal functions or call external functions imported from JavaScript. 
We determine the exact function referring to the interoperations identified in Phase 1.

We now describe our \textbf{data} abstraction rules.
The data abstraction rules characterize suspicious values hidden in the WebAssembly linear memory. 
For example, in Fig \ref{fig:motivation}.c, the malicious payload is stored in the linear memory and loaded when the function \textit{foo} executes.
WebAssembly linear memory is related to two key sections: 
the memory section and the data section, where the former configures the linear memory and the latter initializes it. 
Besides static initialization, the linear memory can also be manipulated at runtime through specific instructions such as \textit{store}.
However, capturing the exact memory through simulation is unaffordable because precise simulation of all the memory manipulation consumes high overhead.
Thus, we limit the scope of data abstraction to the initialization of linear memory.
To be specific, we analyze all of the segments in WebAssembly data sections and abstract them to ``VariableDeclaration'' syntax units as shown in the last row of Table \ref{fig:abstraction}, which means every individual data segment is declared as a unique global variable. 

Finally, both the code and data abstractions are presented as JavaScript-like AST nodes, facilitating seamless integration into the JavaScript PDG.

\textbf{IPDG Integration and JavaScript Reconstruction.}
Once \system finishes abstracting identified WebAssembly units. 
The next step is to integrate the WebAssembly abstractions into the JavaScript PDG so that we can obtain an inter-language PDG. 
The integration takes two kinds of inputs: interoperation positions identified in Phase 1 and the abstraction results generated in Phase 2.
Specifically, \system replaces the PDG nodes identified as invocation positions of WebAssembly functions with their code abstraction, thereby unveiling the previously hidden semantics.
Concurrently, JWBinder integrates the data abstraction of WebAssembly units into their instantiation positions, signifying the initialization of their linear memories.

At present, the IPDG is directly reconstructed into a pure JavaScript program using Escodegen \cite{escodegen}, making it compatible with existing monolingual anti-virus systems. Escodegen, which is a well-tested tool widely adopted in previous work \cite{wobfuscator,escodegen2,hidenoseek}, can generate JavaScript programs from abstract representations following ES6 standards. Nevertheless, we believe that the IPDG structure could potentially serve as a foundation for developing advanced multilingual analysis techniques. We aim to explore this avenue in our future work.
\vspace{-0.2cm}

 % three phases
 \vspace{-0.4cm}
\section{Evaluation}
\label{sec:evaluation}
\vspace{-0.2cm}
Using our \system implementation in Seciton \ref{sec:methodology}, the evaluation of our approach is guided by four research questions below:

\textbf{RQ1:} How effective is \system when deployed with real-world anti-virus solutions?

\textbf{RQ2:} Does \system introduces any side effects to the benign JavaScript-WebAssembly multilingual programs?

\textbf{RQ3:} What is the generalization ability of JWBinder on different commercial anti-virus solutions?

\textbf{RQ4:} How efficient is \system in terms of its runtime overhead.
\vspace{-0.3cm}
\subsection{Experiment Setup and Preliminary Study}
\label{subsec: setup}
We first describe our experiment settings and conduct a preliminary study to illustrate the weakness of existing works against JWMM.

% How to generate the benchmark
\textbf{1) JWMM Dataset.}
%\LPY{We should add a period.}
To evaluate the effectiveness and efficiency of \system, a multilingual malware dataset is required. 
However, to our best knowledge, such a dataset is not available. 
Meanwhile, it may not be feasible to 
curate the ground truth for large complex real-world programs.
Thus, we take inspiration from prior works \cite{li2022polycruise, wobfuscator} to construct a JWMM dataset (JWBench) deriving from real-world JavaScript malware. 
Below, we detail our dataset construction:
\begin{itemize}
    \item Step-1: \textbf{Initial malware selection and filtering}. 
    The initial malware dataset consists of 44369 samples, with 39,450 samples from the Hynek Petrak JavaScript malware collection \cite{hynek}, 3562 samples from VirusTotal\cite{virustotal}, and 1357 samples from the GeeksOnSecurity\cite{gos}.    
    To ensure the semantic validity of the malware samples, We use Esprima \cite{esprima}, a popular standard-compliant JavaScript parser, to vet the initial samples.
    Additionally, we filter the samples which rely on \textit{cc\_on} statements to perform malicious behaviors. 
    \textit{Cc\_on} statement is a special mechanism that only works in IE browser \cite{ie}, which generates executable comments. 
    Thus the detection of such samples is beyond our research scope.
    The final JavaScript malware dataset consists of 21191 JavaScript samples, termed JWBench$_o$. 
    We believe it is of more validity for reasonable evaluation.
    % \item Step-2: \textbf{Malware categories identification}. 
    % The initial malware dataset is blind to specific malware categories. To make our evaluation comprehensive, we upload these samples to VirusTotal and select the report categories with most engines voting. We list the breakdown of the
    % malicious dataset in Appendix.
    \item Step-2: \textbf{JWMM generation}. 
    We employed the technique proposed by Alan et al. in \cite{wobfuscator} to convert real-world JavaScript malware to JWMM for evasion detection. 
    We implement Wobfuscator and apply it to malicious samples from step 1, following the original experiment configuration in \cite{wobfuscator}. As a result,
    the final JWBench contains 21191 JWMM samples.

\end{itemize}
% size, class of malware samples

\textbf{2) Baseline Anti-virus Solutions.} 

To perform a large-scale, representative experiment, we leverage VirusTotal as our baseline anti-virus solution. 
VirusTotal is an online platform housing 74 real-world commercial AV-Systems, including renowned ones like McAfee \cite{mcafee}, Microsoft \cite{microsoft}, and BitDefender \cite{bitdefender}, which makes it the chief target for comprehensive evaluations.
Currently, there are 59 AV-Systems providing detection services for malicious JavaScript. 
When provided a JavaScript program as input, VirusTotal generates a detection report containing binary results (malicious or benign) from these AV-Systems.

% analyze the result
We employed two overall metrics to assess the performance of the AV-Systems on VirusTotal:

\textit{Successful Detection Rate}. A sample is considered successfully detected if no less than a certain threshold number of AV-Systems on VirusTotal accurately identify it as malicious. 
We set this threshold at two to avoid false positives detection referring to \cite{vlabel}. 
The Successful Detection Rate (\textbf{SDR}) signifies the ratio of successfully detected samples to the overall samples.

\textit{Average Detected Engines}.
From a defensive perspective, it is desirable for as many AV-Systems as possible to successfully detect a malicious sample. To this end, we introduce the metric Average Detected Engines (\textbf{ADE}) as another metric to evaluate the capability of AV-Systems on VirusTotal in handling JWMM. The ADE represents the mean number of AV-Systems that successfully detect each malicious sample. 

\textbf{3) Performance of Existing Anti-virus Solutions.}
At a high level, JWMM consists of JavaScript and WebAssembly units. 
However, existing anti-virus solutions might address JWMM from a monolingual perspective without completely considering the interoperations between JavaScript and WebAssembly.
This oversight substantially undermines the detection capability of these anti-virus solutions when dealing with JWMM. 
\begin{table}[]
\vspace{-0.4cm}
    \centering
    \caption{The detection result of VirusTotal on different datasets}
    \begin{tabular}{ccc}
    \toprule
         & JWBench$_o$ & JWBench \\
    \midrule
     Successful Detection Rate & 99.9\%& 47.6\% \\
     Average Detected Engines & 22.4 & 3.3\\
     \bottomrule
    \end{tabular}
    \label{tab:baseline}
    \vspace{-0.5cm}
\end{table}

In Table \ref{tab:baseline}, we list the comparison of VirusTotal's detection results on Different Datasets. As seen in Table \ref{tab:baseline}, on JWBench$_o$, the SDR and ADE of VirusTotal is 99.9\% and 22.4, showcasing its effectiveness against pure JavaScript malware. However, these metrics drop to 47.6\% and 3.3 on JWBench, meaning that more than half of the JWMM samples can elude detection by VirusTotal. 
This result underscores the real-world security threats posed by JWMM.

We further investigate whether anti-virus solutions that target the malicious WebAssembly can detect JWMM effectively. 
Specifically, we first utilize \system's data-flow analysis component to extract 10,000 WebAssembly binaries from JWBench. 
Subsequently, we employ MinerRay\cite{romano2020minerray}, a proven static detector for malicious WebAssembly (e.g., CryptoMiners), to scrutinize these extracted binaries. 
However, MinerRay fails to detect any suspicious activities in all the samples. This result underscores the notion that even specialized anti-virus solutions, despite their deep semantic understanding of WebAssembly, struggle to detect JWMM. This is predominantly because their focus remains limited to heuristic monolingual features.
% VirusTotal

\vspace{-0.4cm}
\subsection{Effectiveness of JWBinder (RQ1)}
\vspace{-0.2cm}
The effectiveness of the \system is evaluated through a comparison of the AV-Systems’ performance on the original JWMM input programs against their performance after \system has been applied. 
For this evaluation, we randomly select 10,000 samples from the JWBench and measure the system's performance using the metrics outlined in \ref{subsec: setup}. 
The primary objectives of \system are to enhance SDR and the ADE of VirusTotal against the selected JWMM samples.

Table \ref{tab:effectiveness} shows the successful detection rate and average detected engines of VirusTotal with/without the application of \system. 
The first column gives the detection  results of VirusTotal without \system, which serves as a baseline.
The rest of the column shows the results after applying \system with different levels of abstraction in semantic reconstruction.

JWBinder$_c$ refers for \system which merely applies code abstraction in SSR and JWBinder$_d$ refers for \system which merely applies data abstraction in SSR. 
The fourth column, JWBinder$_a$, corresponds to the complete version of \system which combines the results of individual JWBinder$_c$ and JWBinder$_d$. 
The values in brackets indicate the difference from the baseline.
For example, the second column indicates that JWBinder$_c$ can successfully detect 16.1\% more malicious samples compared to the baseline, while failing to detect 0.5\% of the samples that the baseline can originally detect.

\begin{table}[]
\vspace{-0.5cm}
    \centering
    \caption{The detection result of VirusTotal on 10k JWMM processed by \system }
    \begin{tabular}{ccccc}
    \toprule
      & Baseline & JWBinder$_c$ & JWBinder$_d$ & \textbf{JWBinder$_a$}\\
    \midrule
     SDR  & 49.1\% & 64.7\% ($\frac{+16.1\%}{-0.5\%}$) & 81.0\% ($\frac{+33.5\%}{-1.6\%}$)& \textbf{86.2\%}($\frac{+37.1\%}{-0.0\%}$)\\
     ADE & 4.1 & 5.0  & 7.5 & \textbf{8.3} \\
     \bottomrule
    \end{tabular}
    \label{tab:effectiveness}
    \vspace{-0.5cm}
\end{table}

Table \ref{tab:effectiveness} illustrates that all variants of \system substantially improve the performance of AV-Systems on VirusTotal in terms of both SDR and ADE.
Specially, JWBinder$_c$ and JWBinder$_d$ achieve 15.6\%/31.9\% increment for SDR and are successful in having each malicious sample detected by 0.9/3.4 additional engines, respectively. 
Moreover, the complete JWBinder$_a$ attains an SDR of 86.2\% and an ADE of 8.3.

Of all the malicious samples, either JWBinder$_c$ or JWBinder$_d$ successfully identifies 59.5\% of them. There are also 5.2\% and 21.5\% unique samples detectable exclusively by JWBinder$_c$ and JWBinder$_d$, respectively. This outcome serves as an ablation study, demonstrating that different SSR levels contribute to the detection of various samples.

The SSR applied by \system could disrupt detection results if it provides AV-Systems with information that may not favor their detection capabilities. For instance, signature-based detection does not respond significantly to code-level information. We attribute the slight decrease in SDR for JWBinder$_c$ (-0.5\%) and JWBinder$_d$ (-1.6\%) to this factor. 
However, by merging the results from both individual SSR applications, \system can enhance detection without any negative impact (i.e., 0\% decrement).

\vspace{-0.5cm}
\subsection{Side Effects (RQ2)}

While \system has demonstrated significant improvements in JWMM detection, it is essential to evaluate its potential side effects. 
Specifically, we need to ensure that \system enhances the effectiveness of AV-Systems on VirusTotal without introducing suspicious behaviors to the benign samples under detection.

A false positive detection occurs when the benign program is wrongly identified as malware by more than one engine.
For RQ2, we evaluate the number of false positive results of VirusTotal when runs on benign samples processed by \system.
Following the experiment settings in \ref{subsec: setup}, we generate a benign JavaScript-WebAssembly multilingual programs dataset from JS150k \cite{js150k}, which contains 150000 JavaScript source files.   
Due to the scaling issues, we randomly select 1000 of them which VirusTotal deems benign and manually confirm the detection results. 
To evaluate whether \system introduces side effects on the samples it processes, we compare the AV-Systems' false positive detections on the original benign input programs when applying \system. 

Our experimental results reveal a relatively low false positive rate introduced by \system. 
Of the 1000 benign samples, JWBinder$_c$ and JWBinder$_d$ only cause 4 and 1 false positive detections, respectively. 
As a result, the JWBinder$_a$

registers a false positive rate of only 0.5\%. 
Given this relatively low false positive rate compared to the considerable enhancements in JWMM detection, we conclude that \system introduces minimal side effects when processing JavaScript-WebAssembly multilingual programs.
\vspace{-0.5cm}

\subsection{Generalization Ability of JWBinder (RQ3)}
\vspace{-0.2cm}

In this research question, we evaluate the generalization ability of JWBinder across different commercial AV-Systems on VirusTotal.
Furthermore, we aim to deduce the internal mechanisms used in these AV-Systems based on the distribution of results by comparing their benefits from different variants of \system.

Table \ref{tab:detection distribution} shows the detection results of different commercial AV-Systems. 
Due to the limitation of length, we only list 5 representative AV-Systems on VirusTotal (the complete table is deferred to Table \ref{tab:full detection distribution} in the Appendix). 
The results show that different levels of SSR benefit certain detectors more rather than others. 
For example, JWBinder$_c$ and JWBinder$_d$ both help Google and Cyren achieve 20\%+ SDR increment, from which we can deduce that these two AV-Systems are likely to classify malicious JavaScript according to either code-level features and data-level features. 
As a result, JWBinder$_a$ has increased the SDR of Google and Cyren from 31.1\%/29.4\% to 61.3\%/55.9\%.

\begin{table}[]
    \vspace{-0.5cm}
    \centering
    \caption{The successful detection rate of individual AV-System on JWMM processed by \system, highest SDR in \textbf{BOLD}}
    \begin{tabular}{ccccc}
    \toprule
      AV-System & Baseline & JWBinder$_c$ & JWBinder$_d$ & JWBinder$_a$\\
    \midrule
      Google & 31.1\% & 53.3\% & 59.4\% & \textbf{61.3\%}\\
      Cyren  &29.4\% & 50.5\% & 53.4\% & \textbf{55.9\%}\\
      McAfee-GW-Edition & 0\% & 47.5\% & 2.4\% & \textbf{47.7\%}\\
      BitDefender              & 14.2\% & 19.7\% & 39.5\% & \textbf{39.7\%} \\
      Microsoft    & 19.3\% & 16.0\% & 35.0\% & \textbf{45.7\%}\\
     \bottomrule
    \end{tabular}
    \label{tab:detection distribution}
    \vspace{-0.5cm}
\end{table}
Some AV-Systems favor particular features more than others. 
For example, with JWBinder$_c$ revealing the code-level features of JWMM, McAfee-GW-Edition's SDR has increased from 0\% to 47.5\%. 
However, the increment is merely 2.4\% with JWBinder$_d$, which shows that McAfee-GW-Edition is relatively insensitive to data-level features.
In Table \ref{tab:detection distribution}, the majority of AV-Systems (4 of 5) gain significantly from JWBinder$_d$, which corresponds to previous research \cite{quarta2018toward} that signature-based matching is wide-adopted in AV-Systems.

\vspace{-0.2cm}
\subsection{Efficiency of JWBinder (RQ4)}
Lastly, we evaluate the run-time performance of \system.
Since \system runs the analysis of each JWMM on a single core, the reported runtime corresponds to a single CPU.  
There are two most time-consuming steps of \system: the data-flow analysis on the JavaScript side and the SSR process which parses and abstracts the WebAssembly binary.
Usually, a JWMM file has a much more complex AST than a traditional JavaScript file, making it time-consuming to be analyzed either dynamically or statically. 
Alan et al. \cite{wobfuscator} shows that the execution time JWMM may increase at most 2079.21\% with 363.52\% larger size compared to traditional JavaScript programs. 
On average, \system needs 10.0 seconds for data-flow analysis to capture the interoperations between JavaScript and WebAssembly. Besides, it needs on average 15.6 seconds for SSR.
Specially, the corresponding median times are 0.7 seconds and 6.7 seconds, while the maximum amounts of time are 207.5 seconds and 202.2 seconds, predominantly when \system processes exceptionally large JWMM files ($>$ 1Mb), with the details provided in Figure \ref{fig:box} within the Appendix.
This result is competitive compared to previous works for large-scale malicious JavaScript detection\cite{doublex,odgen}. 

In particular, we could complete the data-flow analysis for 87.8\% of our JWMM set in less than 10 seconds and SSR for 83.5\% of them in less than 20 seconds. 
This efficiency enables \system to effectively augment existing AV-Systems for detecting JWMM from the wild.
\vspace{-0.5cm}

 % Evaluation
 %\input{tex/RelatedWork} % 
  \section{Limitation}
\label{sec:limitation}

Our current implementation of \system effectively enhances the performance of existing anti-virus solutions. However, it also has a few limitations:
\begin{itemize}
    \item \textbf{Threat of Run-time Code/Data Generation}. 
    \system aims to enhance existing anti-virus solutions during the static analysis phase. Therefore, it is insensitive to run-time behaviors. For example, attackers can dynamically construct malicious payloads in WebAssembly to evade our static data abstractions or generate code at runtime which is beyond the scope of our reconstructed JavaScript. 
    However, none of the existing static approaches can effectively solve this problem, which can only rely on hybrid approaches.
    \item \textbf{Obfuscation}. 
    Existing JavaScript obfuscation techniques may break the semantics required for taint analysis.
    Currently, \system is not equipped with a de-obfuscation component. 
    However, this threat can be mitigated with on-the-shelf de-obfuscation tools such as \cite{jsdeob,de4js}.
    
\end{itemize}
\vspace{-0.5cm} % 
  \section{Conclusion}
\label{sec:conclusion}

In this paper, we propose \system, the first technique for enhancing the detection of JavaScript-WebAssembly multilingual malware (JWMM). 
\system captures the interoperations between JavaScript and WebAssembly and then reconstructs a statically equal JavaScript program, which characterizes the hidden malicious behaviors of JWMM through a novel uniform structure called Inter-language Program Dependency Graph. 
Our evaluation shows the reconstruction process can effectively enhance real-world AV-Systems. 
We also show to what extent can \system benefit Anti-Virus Systems from different vendors.
Finally, we evaluate the efficiency of \system and prove it can be scalable to large JWMM in the real world.  

\section{Acknowledgements}

This work was partly supported by NSFC under No. U1936215 and the Fundamental Research Funds for the Central Universities (Zhejiang University NGICS Platform).
We sincerely appreciate the anonymous reviewers for their insightful comments. 
We also greatly appreciate the authors of Wobfuscator and DoubleX for sharing their code.

\bibliographystyle{splncs04}
\bibliography{tex/bib}

\appendix

\vspace{-0.7cm}
\section{Appendix}
\label{sec:appendix}
\vspace{-0.3cm}
This appendix contains some supplementary material.

\begin{figure}
    \centering
    \scalebox{0.9}{
    \includegraphics[width=\linewidth]{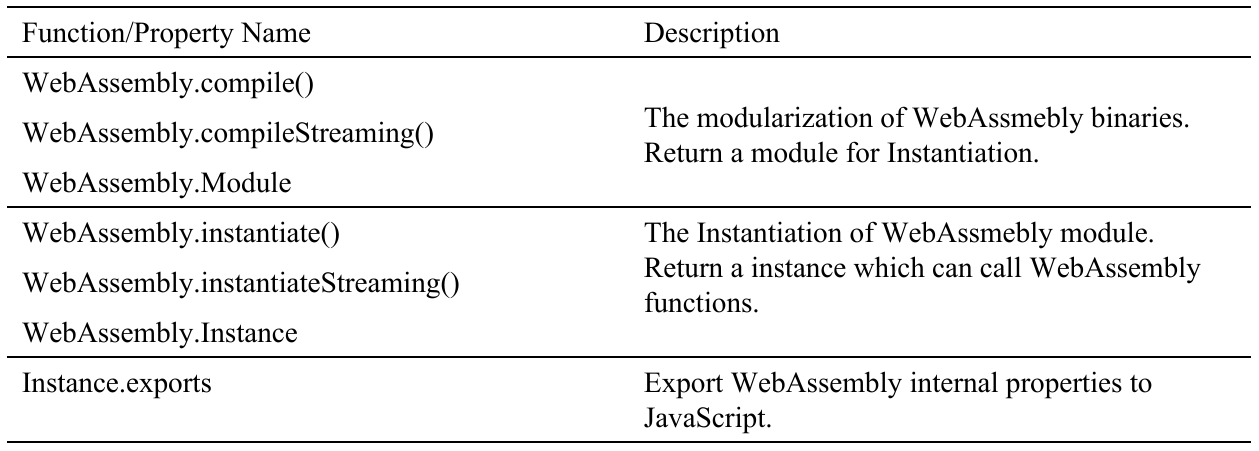}}
    \caption{WebAssembly modularization/instantiation functions and properties}
    \label{fig:keyfunc}
\end{figure}

In particular, Figure \ref{fig:keyfunc} lists a series of JavaScript WebAssembly interfaces for cross-language interoperations, which we leverage in the data-flow analysis.

\begin{figure}[]
    \centering
    \scalebox{0.6}{
    \includegraphics[width=\linewidth]{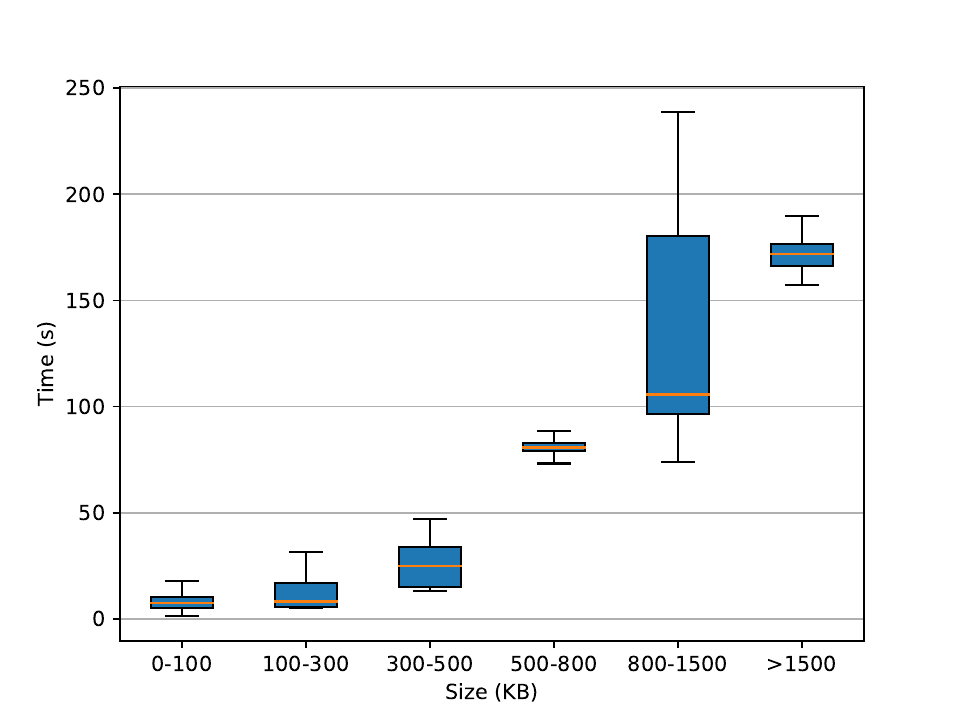}}
    \caption{  Run-time performance of \system depending on the JWMM size}
    \label{fig:box}
\end{figure}

Figure \ref{fig:box} presents \system run-time performance depending on the JWMM size.

\begin{figure}[t]
    \centering
    \includegraphics[width = \linewidth]{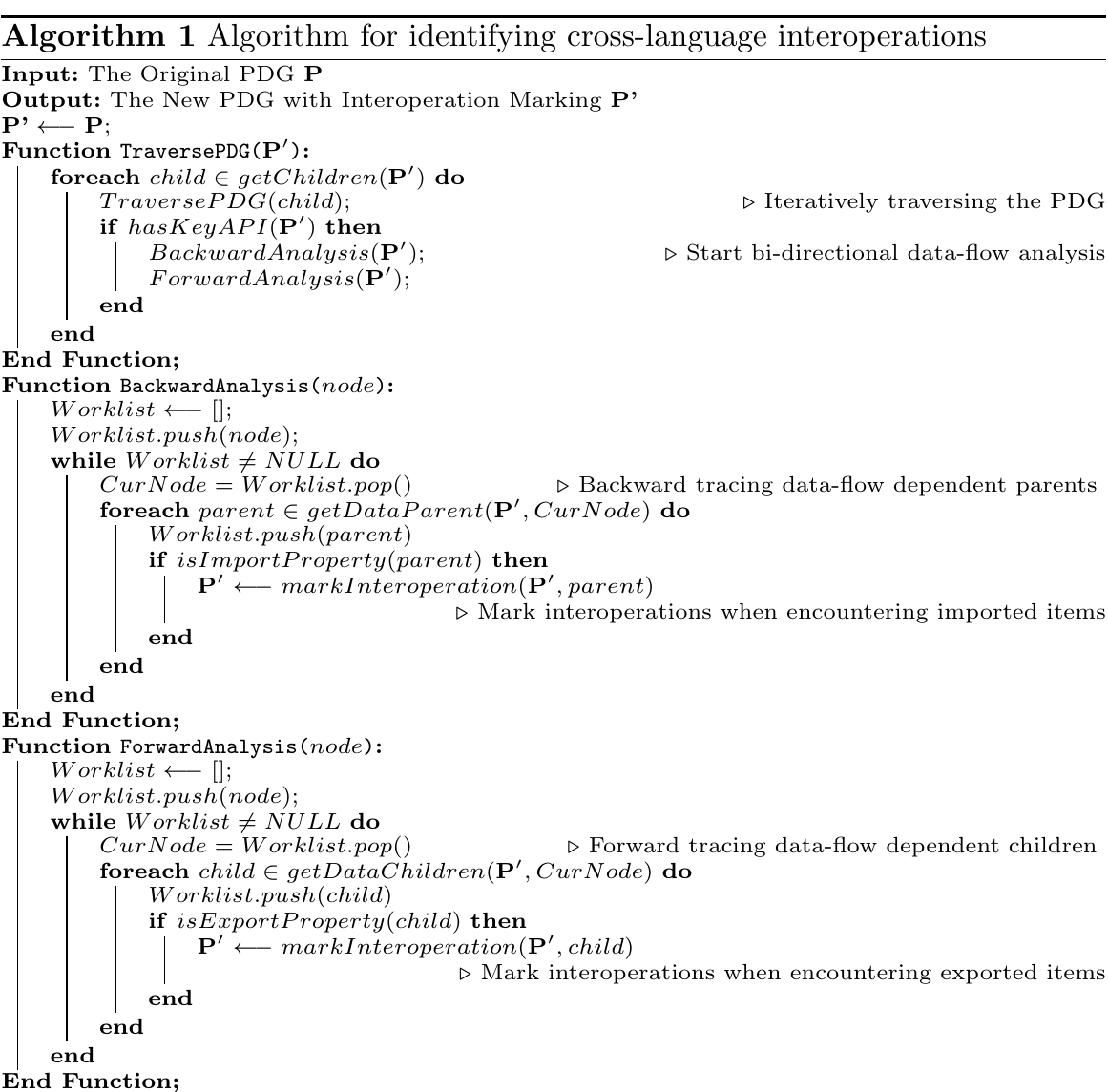}
    \label{fig:algorithm}
\end{figure}

Algorithm \ref{fig:algorithm} details the process for identifying the cross-language interoperations on PDG.

Finally, Table \ref{tab:full detection distribution} shows extensive detection results of different AV-Systems. We list the top 15 AV-Systems due to the space limit.
\vspace{-0.5cm}
\vspace{-1.2cm}

\begin{table}[b]
\vspace{-0.5cm}
    \centering
    \caption{The successful detection rate of every individual AV-System on JWMM processed by \system, highest SDR in \textbf{BOLD}}
    \scalebox{0.9}{
    \begin{tabular}{ccccc}
    \toprule
      AV-System & Baseline & JWBinder$_c$ & JWBinder$_d$ & JWBinder$_a$\\
    \midrule
      Google & 31.1\% & 53.3\% & 59.4\% & \textbf{61.3\%}\\
      Cyren  &29.4\% & 50.5\% & 53.4\% & \textbf{55.9\%}\\
      McAfee-GW-Edition & 0\% & 47.5\% & 2.4\% & \textbf{47.7\%} \\
      Microsoft    & 19.3\% & 16.0\% & 35.0\% & \textbf{45.7\%}\\
      Rising       & 27.5\% & 2.6\% & 44.5\% & \textbf{45.3\%}\\
      Arcabit          & 14.2\% & 20.0\% & 39.8\% & \textbf{39.9\%}\\
      MicroWorld-eScan          & 14.2\% & 20.0\% & 39.8\% & \textbf{40.0\%} \\
      FireEye          & 14.2\% & 19.9\% & 39.5\% & \textbf{39.8\%} \\
      ALYac          & 13.8\% & 19.6\% & 38.9\% & \textbf{39.4\%}\\
      GData          & 14.2\% & 19.0\% & 38.3\% & \textbf{39.3\%} \\
      Emsisoft              & 14.0\% & 18.0\% & 36.9\%& \textbf{38.6\%}\\
      BitDefender              & 14.2\% & 19.7\% & 39.5\% & \textbf{39.7\%}\\
      VIPRE              & 14.1\% & 19.5\% & 39.1\% & \textbf{39.5\%}\\
      MAX              & 14.2\% & 19.9\% & 39.6\% & \textbf{39.8\%} \\
      Ikarus              & 2.07\% & 13.7\% & 23.6\% & \textbf{24.0\%}\\
     \bottomrule
    \end{tabular}}
    \label{tab:full detection distribution}
\end{table}

\end{document}